\documentstyle[epsfig,graphicx]{mn}

\def\etal{{et~al.}}
\def\eg{{e.g.}}
\def\ie{{i.e.}}
\def\cf{{cf.}}
\def\ee{\protect\pee}
\def\pee#1{\ifmmode{\times10^{#1}}\else$\times10^{#1}$\fi}
\makeatletter\def\@cite#1#2{(\if@tempswa #2 \fi #1)}\makeatother
\def\refeq#1{{(\ref{#1})}}
\newcommand{\bv}[1]{\mbox{\boldmath $#1$}}
\def\ion#1#2{{\rm #1}%
\ifmmode{\mathchoice{\scriptstyle}{\scriptstyle}
{\scriptscriptstyle}{\scriptscriptstyle}{\rm\uppercase{#2}}}%
\else$\,\scriptstyle\rm\uppercase{#2}$\fi}
\def\HII{{\ion{H}{II}}}
\mathcode`\|="8000		
{\catcode`\|=\active \gdef|#1{_{\rm #1}}}

\def\unit#1{{\rm\,#1}}
\def\cm{\unit{cm}}
\def\kms{\unit{km\,s^{-1}}}

\title[Magnetic ionization fronts II]{Magnetic ionization fronts II:\\
Jump conditions for oblique magnetization}

\author[R.J.R. Williams, J.E. Dyson \&
T.W. Hartquist]{R.J.R. Williams$^{1,2}$, J.E. Dyson$^2$ and
T.W. Hartquist$^2$\\
$^1$Department of Physics and Astronomy, Cardiff University, PO Box 913, 
Cardiff CF24 3YB\\
$^2$Department of Physics and Astronomy, The University, 
Leeds LS2 9JT
}
\date{Received **INSERT**; in original form **INSERT**}
\pagerange{\pageref{firstpage}--\pageref{lastpage}}

\begin{document}
\label{firstpage}
\maketitle

\begin{abstract}
We present the jump conditions for ionization fronts with oblique
magnetic fields.  The standard nomenclature of R- and D-type fronts can
still be applied, but in the case of oblique magnetization there are
fronts of each type about each of the fast- and slow-mode speeds.  As
an ionization front slows, it will drive first a fast- and then a
slow-mode shock into the surrounding medium.  Even for rather weak
upstream magnetic fields, the effect of magnetization on ionization
front evolution can be important.
\end{abstract}


\section{Introduction}

The sequence of evolution of the \HII\ region around a star with a
strong ionising radiation field which turns on rapidly is well known
(Kahn 1954; Goldsworthy 1961; see also Spitzer 1978, Osterbrock 1989
or Dyson \& Williams 1997).  Initially, the ionization front (IF)
between ionized and neutral gas moves outwards at a speed limited by
the supply of ionizing photons, but it begins to decelerate as the
ionizing flux at the front surface is cut by geometrical divergence
and absorption by recombined atoms within the front.  Eventually, when
the speed of the front decreases to roughly twice the sound speed in
the (now highly overpressured) ionized gas, a shock is driven forwards
into the neutral gas ahead of the front.  Before this stage, the front
is referred to as R-type, while subsequently it is referred to as
D-type.  The shock propagates outwards, gradually weakening, until, in
principle, the \HII\ region eventually reaches pressure equilibrium
with its surroundings.

Where the external medium has an ordered magnetic field, the obvious
critical flow speeds are the fast, Alfv\'en and slow speeds rather
than the isothermal sound speed.  Redman \etal~\cite[1998,
hereafter]{redea98} studied IFs with the magnetic field vector in the
plane of the front, and found that the fast-mode speed plays the same
role as the sound speed does in the hydrodynamic case.  In this paper,
we extend their work to treat the case of an IF moving into a medium
in which the magnetic field is oblique to the direction of propagation
of the front (note also we here follow the more conventional usage in
which the magnetic fields are termed parallel or perpendicular with
respect to the front-normal).

Jump conditions for IFs with oblique magnetization have previously
been studied by Lasker~\shortcite{lask66}.  Here we consider a wider
range of upstream magnetic fields, since observations have shown that
higher magnetic fields are found around \HII\ regions than once
thought likely \cite[\eg{}]{rob93,rob95}.  We determine the properties
of the jumps as functions of upstream conditions, using the velocity
of the front as a parameter rather than as the variable for which we
solve.  We use evolutionary conditions to isolate the stable IF
solutions, and verify these conclusions for a simple model of the
internal structure of the fronts and using numerical simulations.  We
find that rather weak parallel magnetic fields can lead to a
substantial decrease in the D-critical (\ie\ photoevaporation)
velocity from dense clumps except where the magnetic field is exactly
parallel to the IF, and also find additional solutions to the jump
conditions in the range of front velocities forbidden by the
hydrodynamical jump conditions, which were not considered by Lasker.

In the following sections, we present the jump conditions for MHD
shocks (Section~\ref{s:jump}), and discuss the regions for which
evolutionary conditions suggest that these solutions are stable
(Section~\ref{s:solns}).  We verify that the evolutionary solutions
are those with resolved internal structures for one simple model for
the internal structure of the fronts (Section~\ref{s:resolv}).  In the
context of these results, we discuss the development of an IF over
time (Section~\ref{s:devel}) and illustrate this development using
numerical models (Section~\ref{s:numeric}).

Finally (in Section~\ref{s:concl}), we summarize our results, and
provide an example of their application to observations of the \HII\
region S106.  Our physical interpretation of the development of MHD
IFs will, we hope, facilitate the future application of these results.

\section{Jump conditions}

\label{s:jump}
We orient axes so that $\hat{\bv{z}}$ is normal to the front, and
(without loss of generality) that the upstream velocity and magnetic
field are in the $(x,z)$ plane.  We use the usual MHD jump conditions:
\begin{eqnarray}
[\rho v_z]&=& 0 \\{}
[\rho v_z^2 + p + B_x^2/8\pi]& = &0 \\{}
[\rho v_z v_x - B_z B_x/4\pi]& = &0 \\{}
[B_z] & = & 0 \\{}
[v_x B_z - v_z B_x] &=&0,
\end{eqnarray}
except that instead of using the energy flux condition, we adopt the
isothermal equation of state $p = \rho c|s^2$ where the sound speed,
$c|s$, increases across the front but is constant on either side of
it.  We use subscripts 1 and 2 to denote upstream and downstream
parameters, respectively, and write $v_x = u_{1,2}$, $v_z = v_{1,2}$,
$c|s = c_{1,2}$ and $B_x = B_{1,2}$.  Hence
\begin{eqnarray}
\rho_1/\rho_2 = v_2/v_1 &\equiv& \delta \label{e:flux}\\
\rho_2 (v_2^2 + c_2^2) + B_2^2/8\pi &=&
\rho_1 (v_1^2 + c_1^2) + B_1^2/8\pi \label{e:zmom} \\
\rho_2 v_2 u_2 - B_z B_2/4\pi &=& \rho_1 v_1 u_1 -
	B_z B_1/4\pi \label{e:xmom}\\
u_2 B_z - v_2 B_2 &=& u_1 B_z - v_1 B_1 \label{e:freeze}
\end{eqnarray}
Equations \refeq{e:flux}, \refeq{e:xmom} and \refeq{e:freeze} give
\begin{equation}
B_2 = {m_1^2 - 2\eta_1 \over \delta m_1^2 - 2\eta_1} B_1\label{e:b2},
\end{equation}
where $m = v/c|s$ and we define $\eta = B_z^2/8\pi\rho c^2$ and $\xi =
B_x^2/8\pi\rho c^2$ (the $z$ and $x$ contributions to the reciprocal
of the plasma beta).  The dependence on the upstream transverse
velocity has disappeared, as expected as a result of frame-invariance.

In equation \refeq{e:zmom}, we substitute with \refeq{e:b2} for $B_2$
and use equation \refeq{e:flux} to eliminate $\rho_2$ and $v_2$ to
find that, so long as $\delta\ne0$ and $\delta m_1^2\ne 2\eta_1$, the
dilution factor $\delta$ is given by the quartic equation \cite[see
also]{lask66}
\begin{eqnarray}
m_1^6 \delta^4 - m_1^4 (1+m_1^2+4\eta_1+\xi_1)\delta^3\nonumber\\\quad
+ m_1^2(\alpha m_1^2 +4\eta_1(1+m_1^2+\eta_1+\xi_1))\delta^2 \nonumber\\ \quad 
+ (\xi_1 m_1^4 - 4\eta_1(\eta_1+m_1^2(\eta_1+\xi_1+\alpha)))\delta\nonumber\\ 
\quad  +4\alpha\eta_1^2 &=& 0,\label{e:poly}
\end{eqnarray}
where $\alpha = (c_2/c_1)^2$ ($100$ is a typical value).  It is easily
verified that this equation has the correct form in the obvious
limiting cases (of isothermal MHD shocks where $\alpha = 1$, and
perpendicular-magnetized IFs where $\eta_1 = 0$, see Paper I).

\section{Solutions}

\label{s:solns}
\begin{figure*}
\epsfxsize = 8cm
\begin{center}
\begin{tabular}{ll}
(a) & (b) \\
\mbox{\epsffile{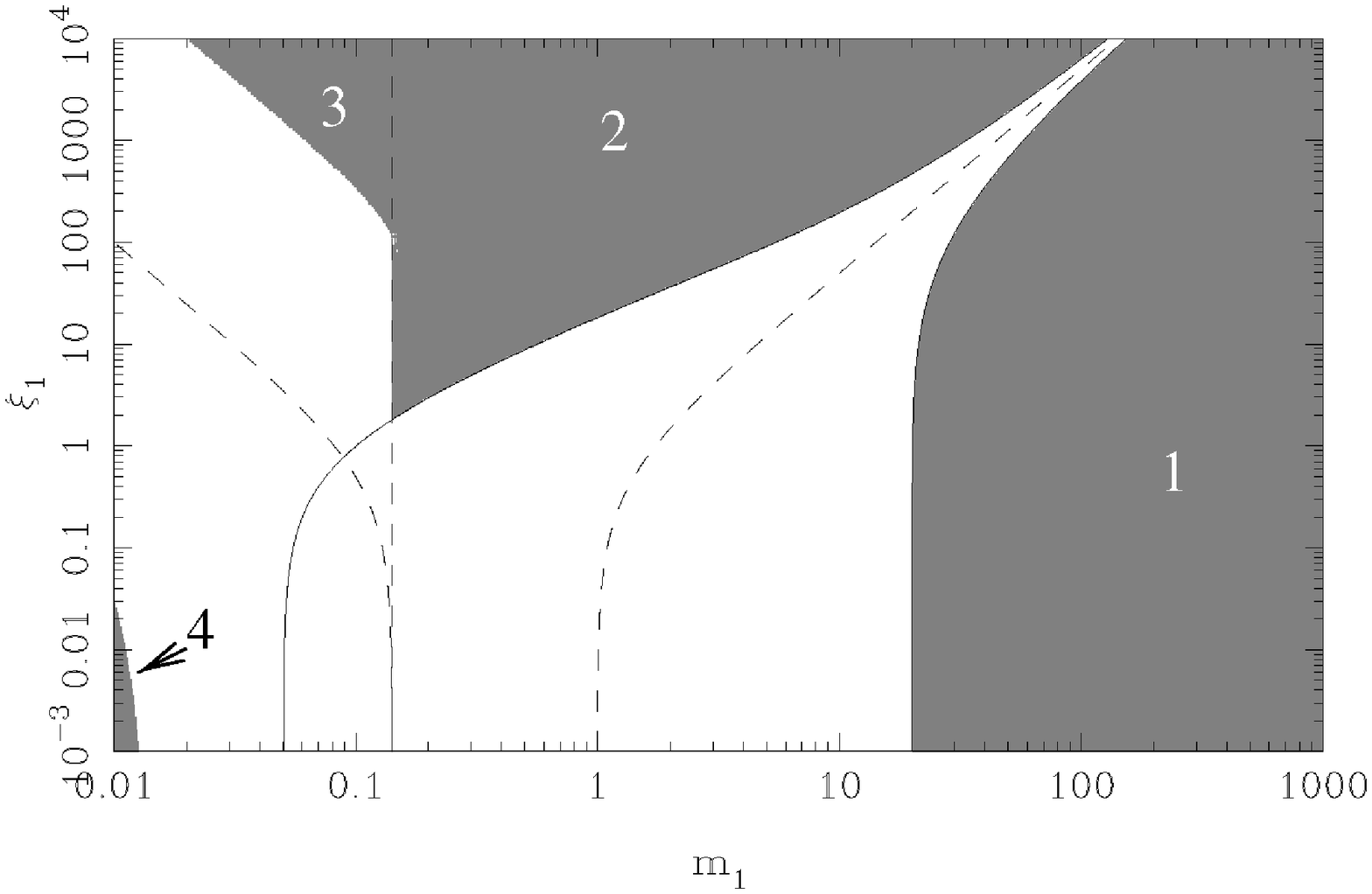}} &
\mbox{\epsffile{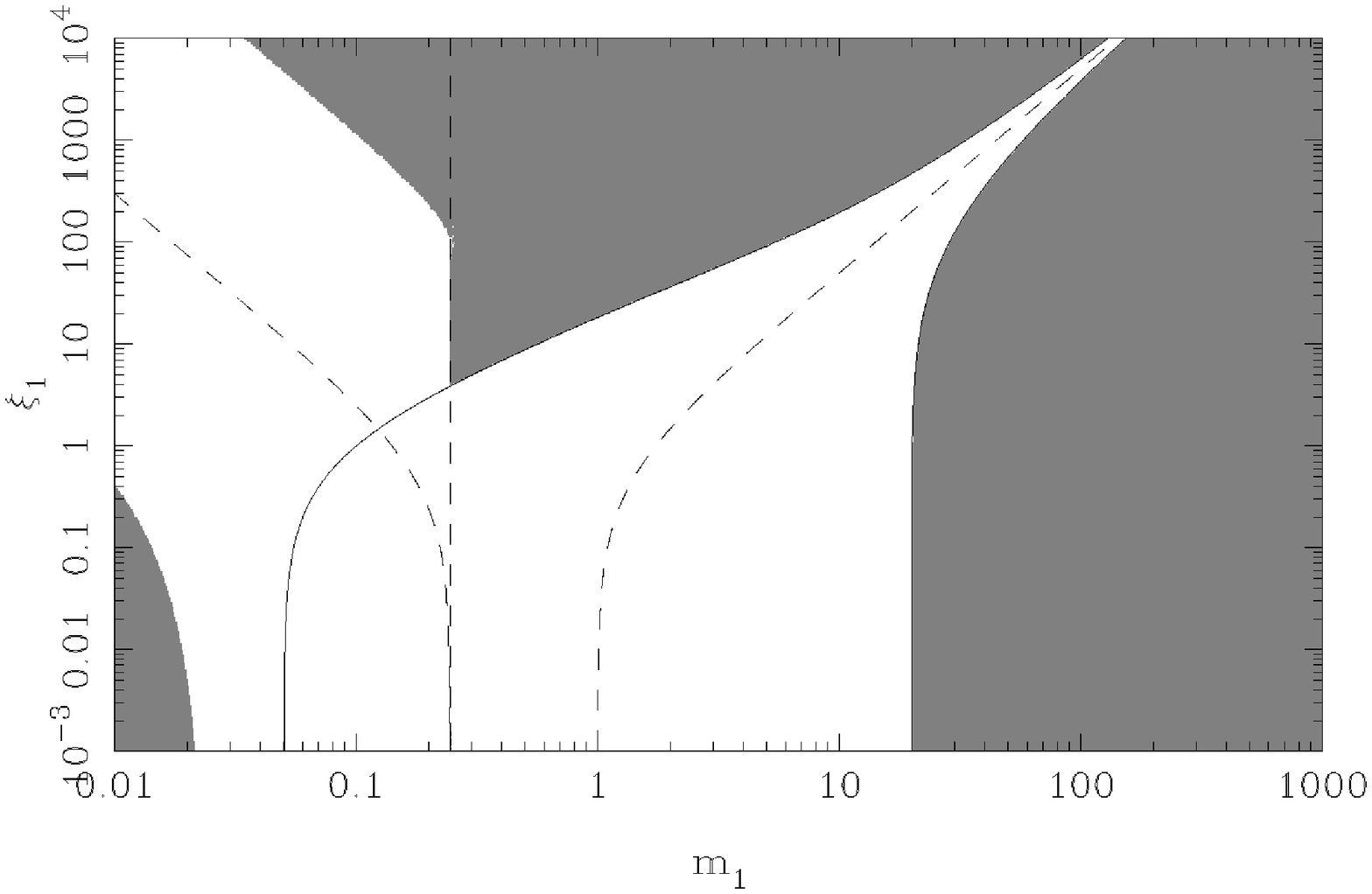}} \\
(c) & (d) \\
\mbox{\epsffile{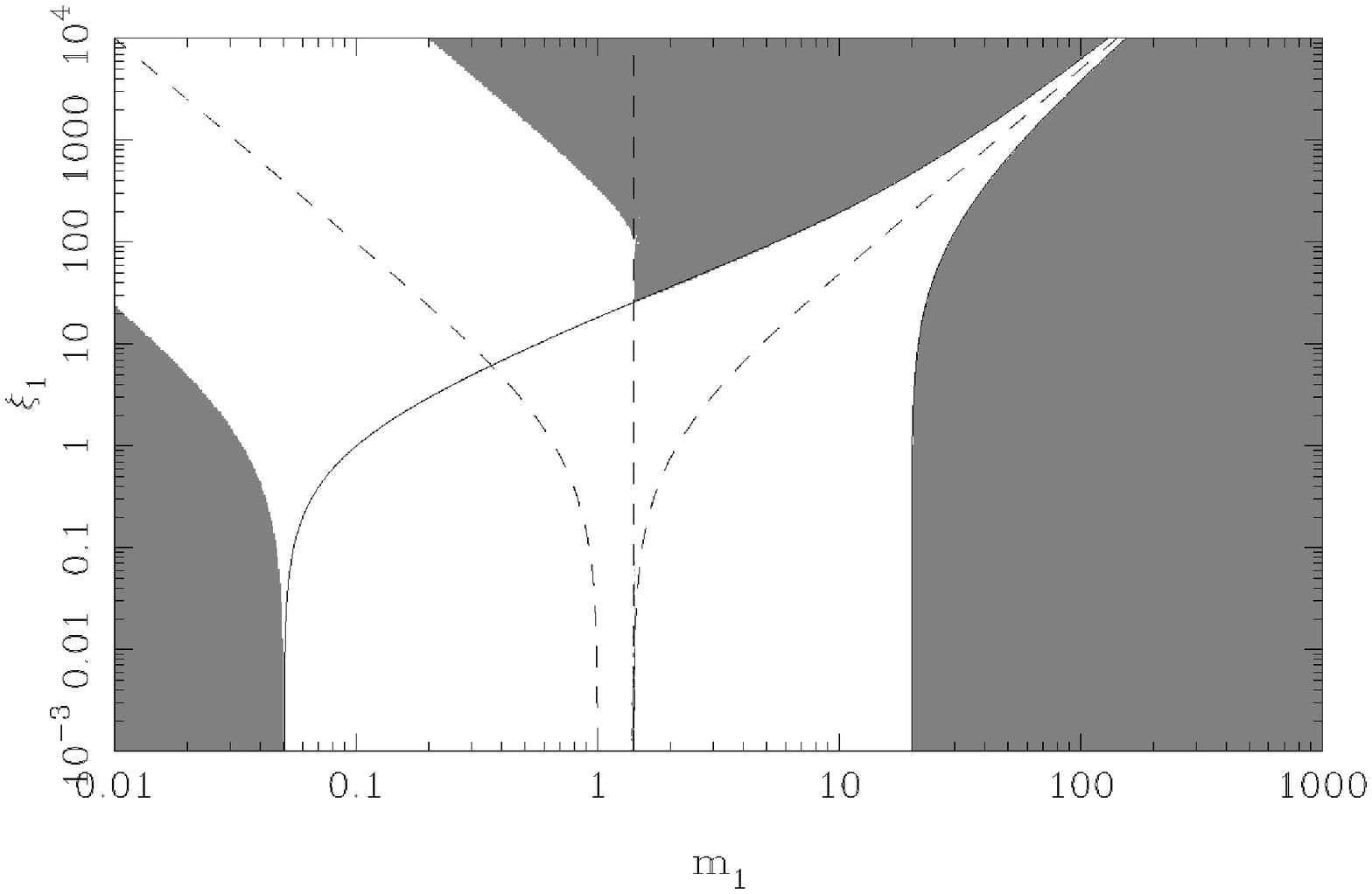}} &
\mbox{\epsffile{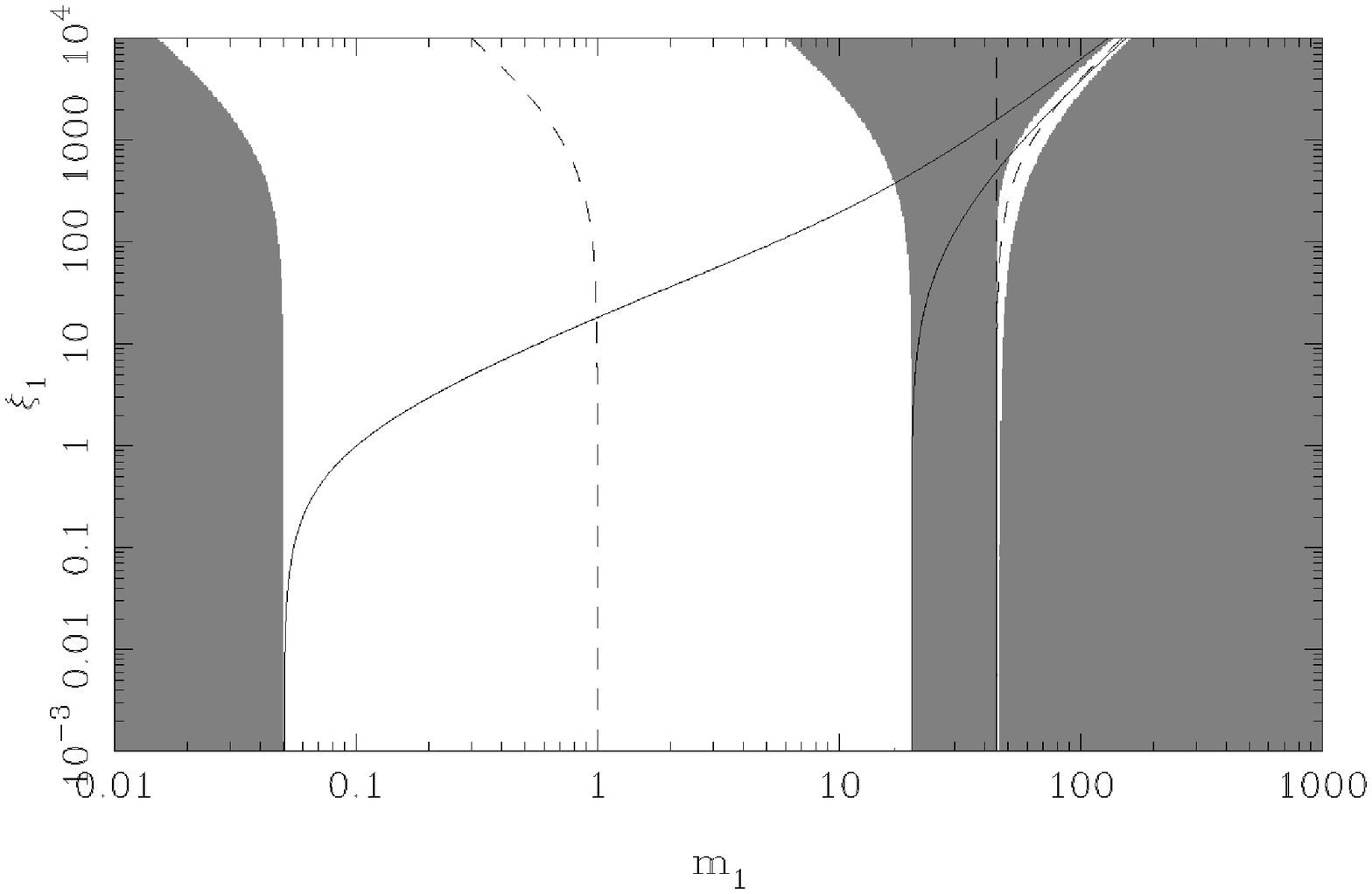}} 
\end{tabular}
\end{center}
\caption{Allowed obliquely magnetized IFs (shown grey).  The solid
lines show the edges of the forbidden region for $\eta = 0$, and the
dashed lines where the inflowing gas moves at the slow, Alfv\'en and
fast speeds, from left to right.  The grey regions correspond to
allowed weak solutions for (a) $\eta_1 = 0.01$, (b) $\eta_1 =
3\ee{-2}$, (c) $\eta_1 = 1$, and (d) $\eta_1 = 1000$.  In (a) the
regions are labelled: 1 -- the $1\to1$, fast-R solutions; 2 -- the
$2\to2$, fast-D solutions; 3 -- the $3\to3$, slow-R solutions; 4 --
the $4\to4$, slow-D solutions.  The fast-D and slow-R regions are
separated by the dashed line at the upstream Alfv\'en speed.  Critical
solutions will be expected at the right-most edge of the slow-D type
region (a $4\to 3$ IF) and of the fast-D type region (a $2\to 1$ IF).}
\label{f:crit}
\end{figure*}

\noindent
Equation~\refeq{e:poly} relates the dilution factor, $\delta$, to the
upstream properties of the flow.  Its roots can be found by standard
techniques.  If the flow is to have a unique solution based only on
initial and boundary conditions, then some of these roots must be
excluded.  It is possible to exclude roots on the basis that they
correspond to flows that do not satisfy an evolutionary condition,
analogous to that long used in the study of MHD shocks.  The
evolutionary condition is based on the requirement that the number of
unknowns in the jump conditions matches the number of constraints
applied to the flow \cite[\eg{}]{jt64}.  For MHD shocks, the number of
characteristics entering the shock must be two greater than the number
leaving it (since the number of independent shock equations is equal
to the number of conserved variables and if there are no internal
constraints on the front structure).

The applicability of the evolutionary conditions to MHD shocks has
been the subject of controversy in the recent past
\cite[\eg{}]{bw88,kbw90}, with various authors suggesting that
intermediate shocks (\ie\ shocks which take the flow from super- to
sub-Alfv\'en speeds) may be stable.  However, Falle \&
Komissarov~\shortcite{fk97,fk99} have shown that the non-evolutionary
solutions are only stable when the symmetry is artificially
constrained, so that the magnetic fields ahead of and behind the
shocks are precisely coplanar.  In any cases in which the boundary
conditions differed from this special symmetry, the solutions
including non-evolutionary shocks were found to be unstable.

The equations governing the dynamics across an IF are no longer a
system of hyperbolic conservation laws, since the ionization source
term cannot be neglected on the scale of the front.  It seems
reasonable to apply analogous evolutionary and uniqueness conditions,
but the mathematical proofs for hyperbolic conservation laws with
dissipation \cite{fk99} no longer apply.  The `strong' evolutionary
conditions suggest that for IF the number of incoming characteristics
is equal to the number of outgoing characteristics, since applying the
ionization equation means that there is an additional constraint on
the velocity of the front.

Solutions which are under-specified by the external characteristic
constraints, termed `weakly evolutionary', may occur if there are
internal constraints, as is the case for strong D-type IFs in
hydrodynamics.  Stable weakly evolutionary solutions only occur for
limited classes of upstream parameters which depend on the internal
structure of the fronts.  Where the number of characteristics entering
the front is greater than suggested by the evolutionary conditions,
the solution can be realized as an MHD shock leading or trailing an
evolutionary IF\@.

In the limit in which $\alpha$ tends to unity from above, the phase
change through the IF has no dynamical consequences, and the IF jump
conditions approach those for isothermal MHD shocks.  This can be seen
if we rewrite equation~\refeq{e:poly} as
\begin{eqnarray}
(\delta-1)[m^6 \delta^3 - m^4 (1+4\eta+\xi)\delta^2 \nonumber\\ +
m^2( 4\eta(1+\eta+\xi)-m^2\xi)\delta \nonumber -4\eta^2] \nonumber\\
\hfill= -(\alpha-1)[m^2\delta-2\eta]^2.\label{e:opoly}
\end{eqnarray}
The IF solution which satisfies the strong evolutionary conditions
becomes the trivial ($\delta=1$) solution of the MHD jump condition.
The other solutions to the IF jump conditions become non-evolutionary
or evolutionary isothermal MHD shocks.  The analysis of Falle \&
Komissarov~\shortcite{fk99} rules out the former as physical
solutions.  The latter are treatable as separate discontinuities,
which will propagate away from the IF when the flow is perturbed,
since the coincidence between the speeds of the shock and of the IF
will be broken.  This argument by continuity supports the use of the
evolutionary conditions for IFs.

As a result of this discussion, we will proceed for the present to
isolate solutions in which the number of characteristics entering an
IF is equal to the number leaving it (and discuss in
Section~\ref{s:resolv} the weakly evolutionary solutions for a
simplified model of the internal structure of IFs).  The velocity of
fronts obeying the strong evolutionary conditions must be between the
same critical speeds in the upstream and downstream gas (somewhat
confusingly, the fronts which obey the strong evolutionary conditions
are termed `weak' in the standard nomenclature of detonations and
IFs).  We follow the usual classification of flow speeds relative to
the fast, Alfv\'en and slow mode speeds $1>v|f>2>v|a>3>v|s>4$, so the
allowed fronts are $1\to1$, $2\to2$, $3\to3$ and $4\to4$.  By analogy
with the nomenclature of non-magnetized IF, we call these fast-R,
fast-D, slow-R and slow-D type IF, respectively.

The panels of Fig.~\ref{f:crit} show regions of $m_1$ and $\xi_1$
space corresponding to evolutionary MHD IFs for several values of
$\eta_1$.  In the figures, we see regions corresponding to the two
distinct classes of R- and D-type solutions.  The flows into the
R-type fronts are super-fast or super-slow, while those into the
D-type fronts are sub-fast or sub-slow.  At the edges of the regions
of solutions either the velocity into the front is the Alfv\'en speed
or the exit velocity from the front is equal to a characteristic speed
(\ie\ the fast mode speed at the edge of the fast-R region, etc.).
For comparison, the solid lines on these plots show the edges of the
forbidden region for perpendicular magnetization ($\eta_1 = 0$, see
Paper I): these lines reach $\xi_1 = 0$ at the edges of the forbidden
region for unmagnetized IF, $0.05\la m_1 \la 20$.

Since equation~\refeq{e:poly} is a cubic in $m_1^2$, quadratic in
$\eta_1$ and linear in $\alpha$ and $\xi_1$, there is no simple
analytic form for the boundaries of the regions.  However, certain
critical values can be determined analytically.  For $\eta_1\la
2\alpha$, the slow-R-critical line terminates where it hits the
Alfv\'en speed at $\xi_1 = \alpha-1$, while the position at which the
fast D-critical line terminates is given by
\begin{equation}
(1+2\eta_1+\xi_1)^2 = 8\alpha\eta_1.
\end{equation}
These points are linked, respectively, by steady switch-off and
switch-on shocks to the points on the limiting slow-D and fast-R
critical loci at which these loci hit the axis $\xi_1 = 0$.  The
intercept between the slow-D critical locus and the axis is at
\begin{equation}
m_1^2 = 2\eta_1 {1-2\eta_1\over\alpha-2\eta_1},
\label{e:slowlim}
\end{equation}
for $2\eta_1 \le (\alpha-\sqrt{\alpha^2-\alpha})$, beyond which the
limiting value is $m_1^2 \simeq 1/(4\alpha)$ as for D-critical
hydrodynamic IF\@.  To illustrate the reason for this change in
solution, we rewrite equation~\refeq{e:poly} for $\xi_1 = 0$ as
\begin{equation}
(m_1^2 \delta-2\eta)^2(m_1^2\delta^2-(1+m_1^2)\delta+\alpha) = 0.
\label{e:noperp}
\end{equation}
The flow leaving a front with no upstream perpendicular component of
magnetic field can be either at the Alfv\'en speed or at the velocity
of the corresponding non-magnetized IF\@.  Where the flow is in the
region beyond the edge of the slow-D-critical region shown in
Figure~\ref{f:crit} (a) or (b), the root of equation~\refeq{e:noperp}
for flow out at the Alfv\'en speed is not a real solution, since
satisfying equation~\refeq{e:zmom} would require that $B_2^2 < 0$.

Even for as small a ratio between magnetic and thermal energy upstream
of the front as implied by $\eta_1 = 0.01$, the effect of parallel
magnetization on the slow-D-critical velocity is dramatic.  Only once
$\eta_1 \la 1/(8\alpha)$ (so the Alfv\'en speed in the upstream gas is
below the unmagnetized D-critical speed) does the fast-critical locus
reach $\xi_1 = 0$, so that the parallel magnetization may be ignored.
As $\eta_1$ increases, the vertical line at the Alfv\'en speed moves
across the plot (see Figure~\ref{f:crit}), decreasing the region of
fast-mode IFs and increasing that of slow-mode IFs.  When
$\eta_1\to\infty$ (a very strong parallel magnetic field), the
(slow-mode) forbidden region is identical to that in the unmagnetized
case (independent of $\xi_1$).

If the upstream flow is at the Alfv\'en velocity, $m_1^2 = 2 \eta_1$,
then the physical solution (when due care is taken with the
singularity of equation~\refeq{e:b2}) is often at $\delta = 1$, \ie\
the ionization of the gas does not change the flow density and it
remains at the Alfv\'en speed.  Both classes of D-type front have
$\delta > 1$ (rarefy the gas), while both classes of R-type have
$\delta < 1$ (compress it).  The perpendicular component of the
magnetic field, $B_x$, increases in a fast-R- or slow-D-type IF, while
it decreases in a fast-D- or slow-R-type.

We will now study the internal structure of the fronts for one simple
model.

\section{Resolved fronts}

\label{s:resolv}
Up to now we have assumed, by investigating the jump conditions, that
the processes within the IF take place on scales far smaller than
those of interest for the global flow problem.  In fact, the flow
structure within an ionization front will vary smoothly on scales
comparable to the ionization distance in the neutral gas.  The flow
may take several recombination lengths behind the front to relax to
equilibrium.  Here we describe the internal structure of MHD IF in one
simple approximation, that the temperature of the gas varies smoothly
through the front but the flow obeys the inviscid MHD equations
throughout \cite[as used in the study of hydrodynamic IF structure
by]{axfo61}.  We shall see that, in this approximation, only fronts
obeying the evolutionary conditions can have smooth structures.  For
the structures to be generic, the jumps across them must satisfy the
strong evolutionary conditions, although singular classes of weakly
evolutionary fronts with internal constraints on their flow structures
are also possible.

\begin{figure}
\epsfysize = 8cm
\mbox{\rotatebox{270}{\epsffile{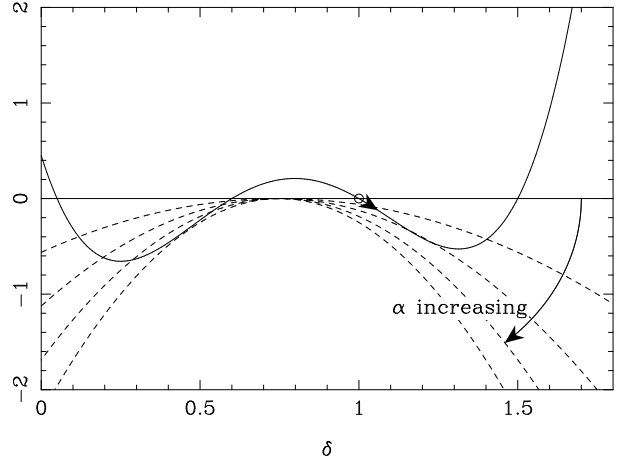}}}
\caption{Schematic plot of equation~\protect\refeq{e:opoly}, where the
solid curve corresponds to the l.h.s. while the dashed curves
correspond to the r.h.s. for various values of $\alpha$.  The points
where the solid curve passes through zero give the solutions of the
isothermal shock jump conditions.}
\label{f:schema}
\end{figure}

In this model the form of an IF is given by the variation of the roots
of equation~\refeq{e:opoly} with the temperature of the gas, specified
by $\alpha$.  The left hand side of this equation is a quartic
independent of $\alpha$, which is positive at $\delta = 0, \infty$ and
zero at $\delta = 1$, and has two or four positive roots \cite[since
the condition that its value is zero is the normal MHD shock condition
for isothermal gas, \cf{}]{ande63}. The right hand side is a
quadratic, which is zero at $m_1^2\delta = 2\eta$ and depends on
$\alpha$ through its (negative) scale.  If the value of $\alpha$
increases steadily through the front, the manner in which the
solutions vary can be followed, as illustrated by the schematic plot,
Fig.~\ref{f:schema}.

For the upstream conditions ($\alpha = 1$), there will be either two
or four solutions where the quartic curve crosses the axis.  Of these,
one is the trivial solution, $\delta = 1$, and at most one corresponds
to an evolutionary shock.  

When internal heating occurs in an IF, the solutions at a given
$\alpha$ correspond to where the solid curve in Fig.~\ref{f:schema}
crosses the corresponding dashed curve,
$-(\alpha-1)(m_1^2\delta-2\eta_1)$.  If gas enters the IF at the point
marked $\circ$, in region 2 between the Alfv\'en speed and the fast
mode speed, then it can move following the arrow as $\alpha$
increases.  For sufficiently large $\alpha$, the dashed curve becomes
tangent to the solid curve: when this occurs, the gas is moving at the
slow- or fast-mode speed, and the IF is called a critical front.  In
between the initial and final solutions, smooth, steady IF solutions
must remain between the same characteristic speeds as they were when
they started.  The strong evolutionary conditions give just those
cases in which a front structure calculated for a smoothly varying,
monotonically increasing $\alpha$ (and non-zero perpendicular magnetic
field) has a continuous solution from the upstream to the downstream
case.

When the upstream flow is at the fast or the slow critical speed, the
l.h.s.\ of equation~\refeq{e:opoly} has a second root $\delta = 1$.
Thus for $\alpha > 1$, this pair of roots disappears, and a forbidden
region is generated.  For $\alpha \le 1$ there is no forbidden region.

For any $\alpha$, the points where the solid curve crosses the dashed
curve in Fig.~\ref{f:schema} are related to each other by the
isothermal MHD shock jump conditions, so a steady shock can form
anywhere within the IF structure for identical upstream and downstream
conditions.  However, since such a flow is over-specified, the equality
of the speed between the shock and IF is a coincidence which will be
broken by any perturbation of the flow (in which case the shock will
escape from one or other side of the IF).  This situation is in direct
analogy with strong R-type IFs in unmagnetized flows.

Strong D-type fronts, for which the exhaust leaves the front rather
above the critical speed, can occur where the heating is not
monotonic.  For these, the highest temperature is attained when the
flow is at the critical speed (\ie\ the curves in
Figure~\ref{f:schema} become tangent exactly at the highest value of
$\alpha$), and as it subsequently cools the solution can move back up
the other branch.  These fronts will form a more restrictive limiting
envelope on the allowed weak solutions than that given by the critical
solutions.  The evolutionary conditions are necessary but not
sufficient, so this behaviour would be expected when more detailed
physics was included.  The actual envelope will correspond to the case
for critical fronts with exhausts at the highest temperature attained
within the front.  Since unmagnetized IF models suggest that any
overshoot in the temperature of the gas is likely to be small, the
envelope will probably not differ greatly from that found for critical
solutions, although the transonic nature of these fronts can be
important in determining the structure of global flows.

Equation~\refeq{e:b2} suggests that no MHD flow can pass through the
Alfv\'en velocity (where $\delta m_1^2 = 2\eta_1$) in a front unless
it has zero perpendicular magnetic field.  Heating the gas will
generally move the flow in regions 2 and 3 away from the Alfv\'en
speed in any case (see Figures~\ref{f:schema} and \ref{f:cusp}).
However, where the perpendicular magnetic field {\it is}\/ zero, a
smooth, weakly evolutionary, front structure can be found (the
internal constraint being zero perpendicular magnetic field).  The
internal structure will be identical to an unmagnetized IF\@.  Indeed,
the strong-D hydrodynamic front can become, by analogy, an
`extra-strong' front which passes through both the Alfv\'en and sound
speeds (for zero perpendicular field, the slow and fast velocities are
each equal to one of these).

By analogy with equation~\refeq{e:b2}, the $y$-components of velocity
and magnetic field are zero everywhere if the MHD equations apply
throughout the front (except if it were to pass through the Alfv\'en
velocity).  Components in these directions {\it can}\/ be generated if
the velocity coupling between different components of the fluid --
electrons, ions, neutrals or dust -- is not perfect \cite[as in shock
structures,]{ph94,ward98}.  These components must, however, damp at
large enough scales: far beyond the front the magnetic field must be
in the same plane and of the same sign as the upstream field, from the
evolutionary conditions.  Exactly this behaviour has been found to
occur in time-dependent multifluid models of MHD shocks (Falle,
private communication).  A full treatment of ionization fronts in
multicomponent material is, however, beyond the scope of the present
paper.

\section{Time development}

\label{s:devel}
In this section, we discuss the development of the IF in a magnetized
\HII\ region, by combining the well-understood development of IF in
unmagnetized environments with the classes of physical roots of
equation~\refeq{e:poly} found above.

An IF driven into finite density gas from a source which turns on
instantaneously will start at a velocity greater than the
fast-R-critical velocity.  Unless the density decreases rapidly away
from the source, the speed of the front decreases so that eventually
the ionized gas exhaust is at the fast-mode speed (at the
fast-R-critical velocity).  When this occurs, two roots of
equation~\refeq{e:poly} merge, and become complex for smaller $m_1$.
As a result, the front will then have to undergo a transition of some
sort.  As in the unmagnetized case \cite{kahn54}, if the size of the
ionized region is large compared to the lengthscales which
characterise the internal structures of shocks and IFs, the IF will
evolve through emitting (one or more) shocks.  The evolution of an
initial IF discontinuity can be treated as a modified Riemann problem
because the speed of the IF is determined by the flow properties on
either side of it, together with the incident ionizing flux which we
assume varies slowly.  The development of this modified Riemann
problem will be self-similar, just as for conventional Riemann
problems.  One complication is that the IF may be located within a
rarefaction wave, but this does not occur for the circumstances we
discuss in the present section.

The simplest possibility for a slowing fast-R-critical IF is that it
will become fast-D-type by emitting a single fast-mode shock.  This
has obvious limits to the cases where the magnetization is zero (where
the shock is a normal hydrodynamic shock), and where it is purely
perpendicular.  If the speed of the IF is specified by the mass flux
through it, then the leading shock driven into the surrounding neutral
gas must be a fast-mode shock, since the upstream neutral gas must
still be advected into the combined structure at more than the
Alfv\'en speed after the transition, and so only a fast-mode shock can
escape.

While there may be no fast-D-type solutions at the value of $\xi_1$
which applied for the fast-R-type front, the fast-mode shock moving
ahead will act to increase the value of $\xi$ upstream of the IF,
since the (squared) increase in the perpendicular component of the
magnetic field dominates over the increase in the gas density after
the shock.  At fast-R-criticality, there is in general a second
solution with the same upstream and downstream states in which a
fast-mode shock leads a fast-D-critical IF, because the l.h.s.\ of
equation~\refeq{e:opoly} (for which a zero value implies an isothermal
shock solution) is positive for large $\delta$ and for post shock flow
at the Alfv\'en speed (\ie\ $m_1^2\delta = 2\eta_1$), unless the flow
into the front is at the Alfv\'en velocity, or $\eta_1$ or $\xi_1$ is
zero.  For example, for a downstream state $\eta_2 = \xi_2 =
5\ee{-3}$, there are two evolutionary solutions: a $1\to1$ front with
$\eta_1 = 0.994$, $\xi_1 = 0.248$, $m_1 = 20.0$ and $\delta=0.503$,
and a $2\to2$ front with $\eta_1 = 3.11\ee{-2}$, $\xi_1 = 11.1$, $m_1
= 0.625$ and $\delta = 16.1$.  Figure~\ref{f:crit} parts (c) and (b),
respectively, contain points which correspond to these solutions.
These two upstream states are linked by a fast-mode shock with the
same velocity as the IF\@.  Thus the emission of a single fast-mode
shock is a valid solution of the modified Riemann problem which occurs
as the flow passes through criticality {\it whatever}\/ the internal
structure of the shock and IF, so long as the evolutionary shocks and
IFs exist.  (Since the l.h.s.\ of equation~\refeq{e:opoly} is greater
than zero so long as $\eta_1\ne0$ for $\delta = 0$, an equivalent
argument holds for slow-critical transitions.)

It is possible that further waves may be emitted at the transition,
for instance a slow-mode shock into the neutral gas together with a
slow-mode rarefaction into the ionized gas.  For the model resolved
IF, this seems unlikely to occur unless the flow velocity reaches the
slow-mode speed somewhere within the fast-D-critical IF\@.  These other
solutions are not seen in our numerical simulations below.  Additional
solutions would also make the development of the IF non-unique, if the
simpler possibility is allowed.

The fast-critical transition may be followed using the jump conditions
for a front with its exhaust at the fast-mode speed (\ie\ a
fast-critical front).  In the smoothly-varying $\alpha$ model of
Section~\ref{s:resolv}, the upstream and downstream states can be
joined by a front in which an isothermal MHD shock is at rest in the
IF frame {\it anywhere}\/ within the IF structure, since the
quantities conserved through the front are also conserved by the
shock.  The evolution of an IF through criticality will occur by an
infinitesimally-weak fast-mode wave at the exhaust of the IF moving
forward through its structure and strengthening until it eventually
escapes into the neutral gas as an independent shock \cite[as
illustrated for recombination fronts by]{wild96}.

The escaping fast-mode shock leads to a near-perpendicular field
configuration upstream of the IF\@.  This boost in the perpendicular
component is required if the transition is to proceed though the
fast-D type solutions, which, as Figure~\ref{f:crit} illustrates, are
near-perpendicular (large $\xi_1$) except where $B_z$ is very large or
very small.  This will result in a rapid change in downstream
parameters across a front where the upstream field is nearly parallel
to the IF\@.  As an example, the flow downstream of the IF will either
converge onto or diverge from lines where the upstream magnetic field
is parallel to the ionization front, and as a result may produce
inhomogeneities in the structure of \HII\ regions on various scales
(from bipolarity to clumps).

\begin{figure}
\epsfxsize = 8cm
\mbox{\epsffile{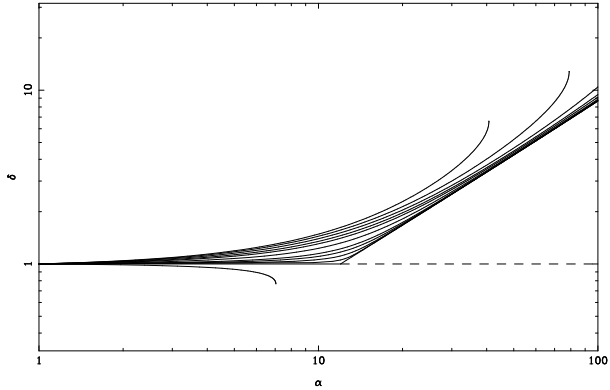}}
\caption{Plot of the dilution factor, $\delta$, within magnetized IFs
for $\eta_1=3\ee{-2}$ and $\xi_1=11$ (\cf\
Figure~\protect\ref{f:crit}(b) and the numerical example in the text),
and $m_1^2$ decreasing towards $2\eta_1$ (with values for $m_1$,
plotted from top to bottom of the fan entering at $\delta=1$, of $m_1
= 1$, 0.7, 0.5, 0.4, 0.35, 0.3, 0.27, 0.26, 0.252, 0.247, the critical
solution and 0.22).  Also shown dashed are the non-evolutionary
Alfv\'en solution for critical $m_1$.  The solutions terminate for
$\alpha < 100$ for both high and low values of $m_1$.  Note also the
development of the gradient discontinuity at $\alpha = \xi_1+1$ in the
density structure.}
\label{f:cusp}
\end{figure}

As the velocity of the front decreases further, eventually it will
approach the Alfv\'en speed.  We find that the evolution depends
qualitatively on whether $\alpha-1$ is smaller than $\xi_1$.  In
Fig.~\ref{f:cusp}, we plot $\delta$ as a function of $\alpha$ for a
range of values of $m_1$ (\ie\ the internal structure of the fronts in
the simple model of Section~\ref{s:resolv}).  We take
$\eta_1=3\ee{-2}$ and $\xi_1 = 11$, corresponding to our numerical
example above, so the ratio of the upstream Alfv\'en speed to the
upstream isothermal sound speed is $0.245$.  For $m_1$ slightly larger
than $0.245$ (\ie\ just super-Alfv\'en), the curves remain flat until
$\alpha$ is close to $\xi_1+1$ and then turn upwards when they reach
this value.  If the value of $\alpha$ in the fully ionized gas were
less than $\xi_1+1$, the solutions would move through a case where the
flow is of uniform density and moves at the Alfv\'en speed throughout
the front before becoming slow-R-critical for some $m_1^2 < 2\eta$, as
can be seen in the leftmost part of Fig.~\ref{f:cusp}, for fronts in
which the maximum $\alpha$ is smaller than 12.

For larger values of $\alpha$, however, the solutions develop a
gradient discontinuity in their structure when the inflow is at the
Alfv\'en speed, $m_1^2 = 2\eta$.  At this discontinuity, the
transverse component of the magnetic field becomes zero (\ie\
switch-off occurs).  Once the propagation speed of the IF drops below
the Alfv\'en speed, there is no form of steady evolutionary structure
with a single wave in addition to the IF which is continuous with that
which applied before.  Non-evolutionary $3\to2$ type solutions do
exist for fronts just below this limit, but in numerical simulations
(see Section~\ref{s:numeric}) these break up.  A slow-mode switch-off
shock moves into the neutral gas and a slow-mode switch-on rarefaction
is advected away into the ionized gas.  Between them, these waves
remove the parallel component of magnetic field at the D-type IF
between them.  A precursor for the rarefaction is apparent in internal
structure of the critical front, Figure~\ref{f:cusp}.  We find in
numerical simulations that the IF which remains is trans-Alfv\'enic.
Note that a steady resolved structure is possible for such (weakly
evolutionary) fronts only because it has exactly zero parallel field
throughout.

Analogous processes must occur in an accelerating {\it fast}\/-mode IF
with weak parallel magnetization when $\eta_1$ is very large: in
Fig.~\ref{f:crit} (d), it is the fast-D-critical line rather than the
slow-R-critical line which meets the Alfv\'en locus at finite $\xi_1$.

\section{Numerical solutions}

\label{s:numeric}
\begin{figure}
\epsfysize = 8cm
\begin{center}
\begin{tabular}{l}
(a)\\
\mbox{\rotatebox{270}{\epsffile{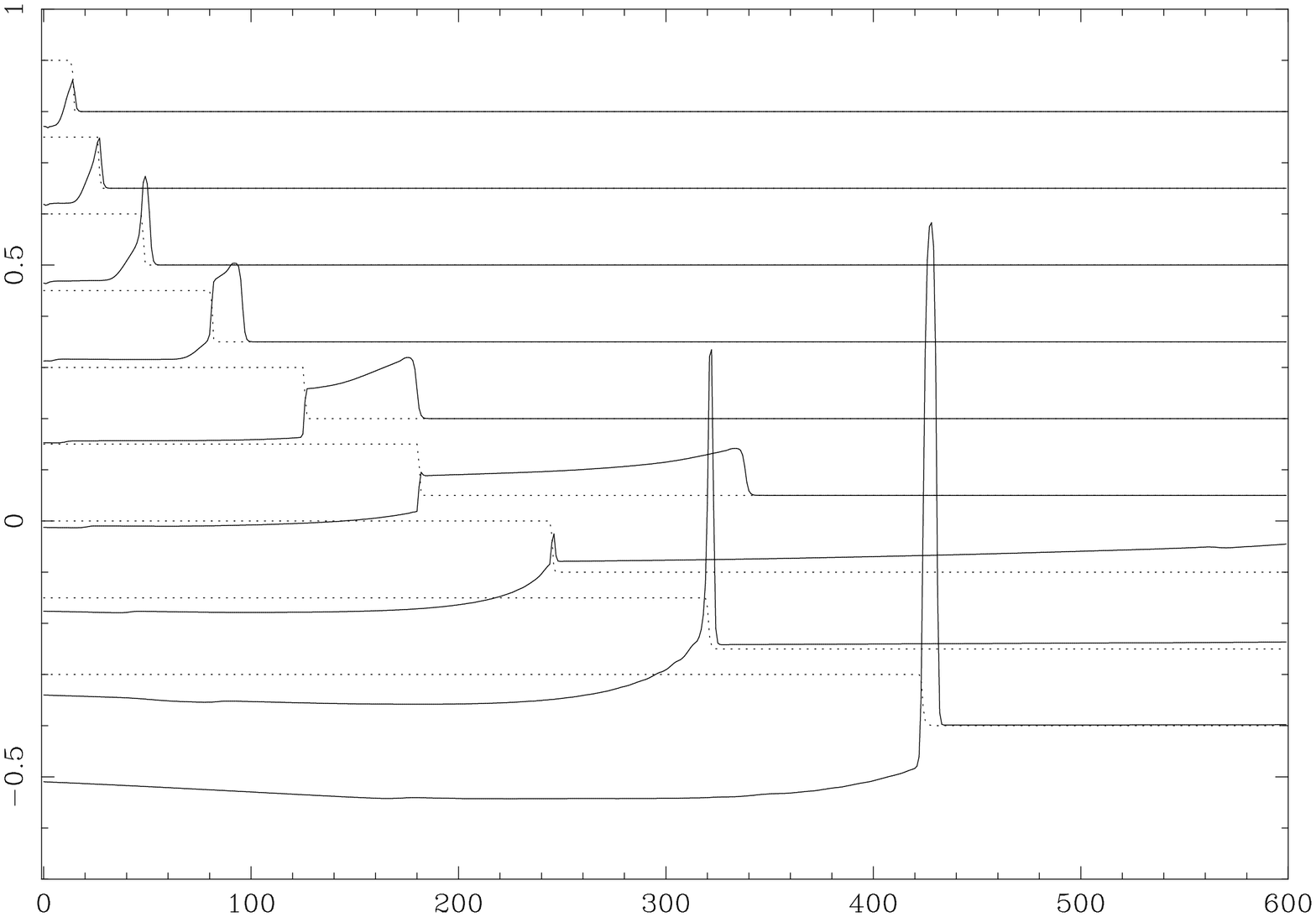}}}\\
(b)\\
\mbox{\rotatebox{270}{\epsffile{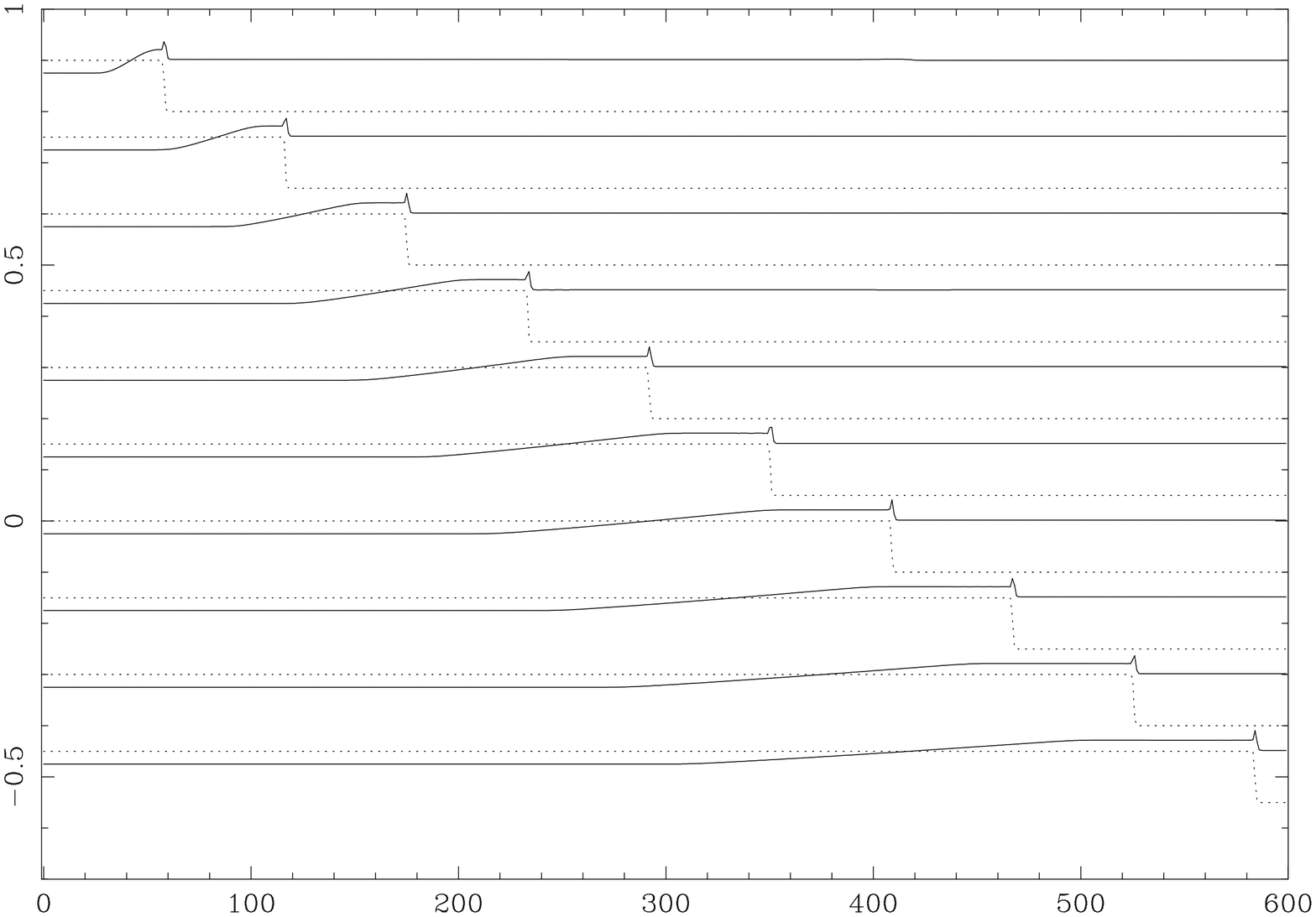}}}
\end{tabular}
\end{center}
\caption{Evolution of an oblique magnetized IF for (a) gradually
decreasing incident radiation intensity; (b) constant intensity chosen
to show slow-R-type IF\@.  In (a), the frames show log base 10 of
density (solid) and ionization fraction (dotted), offset by a constant
between each plot, plotted at times $4,8,16,...,1024$.  The topmost
plot shows a fast-R IF, while third, fourth and fifth show a fast-D IF
together with a fast shock moving to the right.  In the next two, the
IF changes to a slow-R type while in the final two a slow shock has
moved off leaving a slow-D type behind.  In (b), a more clearly
resolved slow-R type front is shown.  In this figure, the density is
plotted linearly, scaled by a factor 0.1, and the plots are at times
$200,400,...,2000$.}
\label{f:evol}
\end{figure}

\begin{figure}
\epsfysize = 8cm
\begin{center}
\mbox{\rotatebox{270}{\epsffile{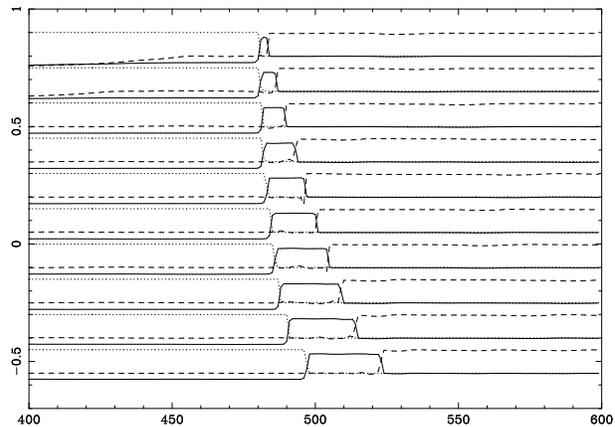}}}
\end{center}
\caption[]{Evolution of an IF from an initially non-evolutionary
solution.  The curves show the density (0.1 times the log base 10,
solid), ionization fraction (multiplied by 0.1, dotted) and
$0.1\times B_x/\sqrt{4\pi}$ (dashed), plotted at times
$400,800,1200,...,4000$ with an offset between each plot.}
\label{f:nevol}
\end{figure}

\begin{figure}
\epsfysize = 8cm
\begin{center}
\mbox{\rotatebox{270}{\epsffile{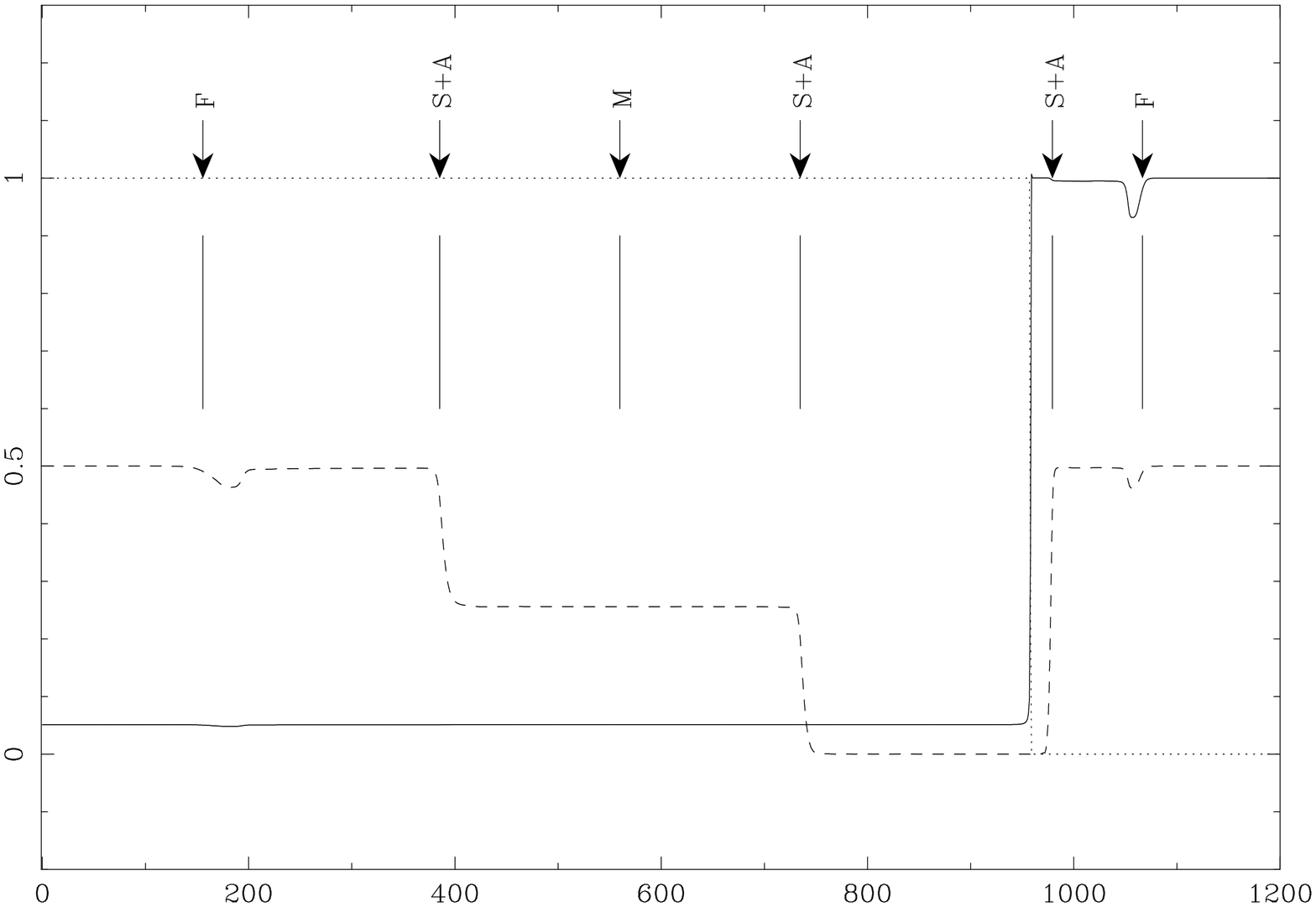}}}
\end{center}
\caption[]{Evolution of an IF obeying the hydrodynamic D-critical
conditions at rest, perturbed by an initial magnetic field.  The
upstream density is $1$ and velocity is $-0.0513$, while the
downstream density is $0.0513$ and velocity is $-1$.  A parallel
field of $B_z = 0.1\sqrt{4\pi}$ means this front is trans-Alfv\'enic.
The IF is perturbed with a perpendicular magnetic field of magnitude
$0.03\sqrt{4\pi}$, in different planes up- and down-stream.  The solid
line shows the density, the dotted line is the ionization fraction and
the dashed line is one component of the perpendicular magnetic field
(which has been scaled to a maximum amplitude of 0.5).  Also shown are
the positions of the characteristic waves from the initial
discontinuity (F is fast, A is Alfv\'en, S is slow, M is mass), based
on the initial states.  The waves which change the magnetic field are,
from left to right, a slow-mode shock plus an Alfv\'en wave, a
slow-mode switch-on rarefaction (which switches {\it off}\/ the field
as it is being advected away from the IF) and a slow-mode switch-off
shock propagating into the upstream gas.  The front which remains is
trans-Alfv\'enic, but has effectively zero perpendicular field.}
\label{f:dcrit}
\end{figure}

In this section, we present some numerical examples to illustrate the
processes discussed in the preceding sections.  We have implemented
linear scheme A as described by Falle, Komissarov \&
Joarder~\shortcite{fkj98} in one dimension, and added an extra
conserved variable corresponding to the flow ionization.  To study the
local development of the IFs we have neglected recombination terms and
just chosen to set the mass flux through the ionization flux as a
function of time.

Figure~\ref{f:evol}(a) shows the propagation of an IF into gas with
density $1$, $B_z = 0.3\sqrt{4\pi}$, $B_x = 2\sqrt{4\pi}$.  We reset
the flow temperature at the end of each time step so that $p =
(0.1+0.9x)\rho$ (the temperature ratio between ionized and neutral gas
is rather small so that any shells of shocked neutral gas are more
easily resolved).  For these values, the characteristic speeds in the
neutral gas are $v|s = 0.046$, $v|a = 0.3$ and $v|f = 2$.  In the
first figure, the incident flux varies as $80/(20+t)$, and the
transitions through fast-R, fast-D and slow-D are clearly visible,
while the slow-R stage (which is often narrow in the parameter space
of Fig.~\ref{f:crit}) is less clear.

In Figure~\ref{f:evol}(b) the upstream conditions are the same, but
the flux was set to a constant value of $0.26$ in order to isolate the
slow-R transition.  This is between the upstream slow-mode and
Alfv\'en speeds, and equation~\refeq{e:poly} predicts that a slow-R
transition exists which will increase the flow density from $1$ to
$1.18$ (for comparison, the $3\to4$ jump relations require the density
to increase to 1.75).  In the simulation, a weak fast-mode wave
propagates off from the IF initially, but does not greatly change the
upstream conditions.  The slow-R IF which follows it increases the
density, and is backed by a rarefaction because of the reflective
boundary condition applied at the left of the grid.  The small
overshoot within the front is presumably due to numerical viscosity,
and can be removed by broadening the IF \cite[\eg\ by the method
described in]{will99}.  The plateau between the rarefaction and the IF
has density $1.22$, which is in adequate agreement with the jump
conditions (particularly when account is taken of the slight
perturbation of the conditions upstream of the front).

These numerical solutions illustrate the orderly progression of a
magnetized IF through the various transitions described in the
previous section.  In a realistic model, however, the density
perturbations generated by the transitions will have important effects
on the evolution, as the consequent changes in recombination rates
alter the flux incident on the IF\@.  These processes should ideally be
studied in the context of a two- or three-dimensional global model for
the evolution of magnetized \HII\ regions, which is beyond the scope
of the present paper.

In Figure~\ref{f:nevol}, we illustrate the development of the flow
from initial conditions in which a non-evolutionary IF is stationary
in the grid.  The upstream (neutral) gas has density $1$, $B_z =
0.3\sqrt{4\pi}$, and $B_x = \sqrt{4\pi}$ and moves into the front at
$v_z = -0.253$ (with no transverse velocity), while the downstream
(fully ionized) gas has $\rho = 0.423$, $v_z = -0.598$, $v_x = 1.688$
$B_z = 0.3\sqrt{4\pi}$, and $B_x = -0.424\sqrt{4\pi}$ so the IF is of
$3\to2$ type.  We set the pressure as above, and the value of the
incident flux as $0.253$ so the initial IF would remain steady in the
grid.  The IF breaks up immediately, driving a slow-mode shock away to
the right, into the neutral gas, while a slow-mode rarefaction moves
away to the left, into the ionized gas.  The D-type IF which remains
is marginally trans-Alfv\'enic, but has zero transverse magnetic
field.

To study this further, in Fig.~\ref{f:dcrit}, we illustrate a
simulation of an IF close to the (hydrodynamic) D-critical condition,
with a parallel magnetic field which makes it trans-Alfv\'enic.  When
perturbed with small but significant perpendicular field components,
this IF again switches off these fields by emitting slow-mode waves.
It remains stable, satisfying the hydrodynamic jump conditions as a
weakly evolutionary solution of the MHD jump conditions.

\section{Conclusions}

\label{s:concl}
We have presented the jump conditions for obliquely-magnetized
ionization fronts.  We have determined the regions of parameter space
in which physical IF solutions occur, and have discussed the nature of
the interconversions between the types of front.  Fast-D and slow-R
solutions with high transverse fields are found in the region of front
velocities forbidden by the hydrodynamic jump conditions: in the
evolution of an \HII\ region, the fast-mode shock sent into the
neutral gas by the fast-critical transition will act to generate these
high transverse fields.  In the obliquely-magnetized case, the fronts
are significantly perturbed as long as the Alfv\'en speed in the
neutral gas is greater than $c_1^2/2c_2$.  However, the stability of
parallel-magnetized weakly evolutionary IF means that the flow may
still leave at the isothermal sound speed in the ionized gas over much
of the surface of magnetized globules exposed to ionizing radiation
fields.

Large ($\ga100\,\mu{\rm G}$), highly ordered magnetic fields have been
observed in the molecular gas surrounding some \HII\ regions
\cite[\eg{}]{rob93,rob95}.  Roberts \etal~\shortcite{rob95} suggested
that the highest observed magnetizations in S106 are associated with
unshocked rather than shocked gas, as a result of the relatively low
density and velocity shift observed for the strongly magnetized gas.
They suggested that the magnetic field becomes tangled close to the
IF, leading to the decrease in detectable magnetization.

It is interesting to compare these results with the example
fast-critical front we discuss in Section~\ref{s:devel}.  Scaling the
parameters of this front to an exhaust hydrogen density of
$10^4\cm^{-3}$, typical of an ultracompact \HII\ region, a mean mass
per hydrogen nucleus of $10^{-24}{\rm\,g}$, and a sound speed in the
ionized gas of $10\kms$, the limiting $1\to1$ front takes the flow
from an upstream state with $n|H = 5.03\ee3\cm^{-3}$, $\bv{v} =
(-0.0497,0,20.0)\kms$ and $\bv{B}=(17.7,0,35.4)\,\mu{\rm G}$ to an
exhaust with $n|H = 10^4\cm^{-3}$, $\bv{v} = (0,0,10.1)\kms$ and
$\bv{B}=(35.4,0,35.4)\,\mu{\rm G}$ (while the fast-R-critical speed is
almost exactly twice the exit speed of the front, this ratio decreases
in more strongly magnetized fronts). The limiting $2\to2$ front takes
the flow from $n|H = 1.61\ee5\cm^{-3}$, $\bv{v} = (1.78,0,0.625)\kms$
and $\bv{B}=(671,0,35.4)\,\mu{\rm G}$ to the same final state.  A
fast-mode shock ahead of the front and at rest with respect to it
would change the upstream state from that of the limiting $1\to1$
front to that of the limiting $2\to2$ front.

This fast-mode shock, which precedes a limiting fast-weak-D type
front, boosts the $x$-component of the magnetic field from
$18\,\mu{\rm G}$ to $670\,\mu{\rm G}$, with a $20\kms$ change in the
$z$-velocity and only a $2\kms$ change to the $x$-velocity component.
A succeeding slow-mode shock would further increase the $z$-velocity
component and gas density while weakening the $x$-component of
magnetic field, without recourse to field-tangling.  If we tentatively
identify OH component B of Roberts \etal\ as fast-shocked material and
component A as doubly-shocked material, component B is more
edge-brightened and has a smaller blueshift than component A as would
be expected.  Component A is kinematically warmer and most blue
shifted towards the centre of the region.  The strong line-of-sight
magnetic fields in S106 are seen at the edges of the region, in a
`toroidal' distribution.  The (poorly resolved) line-of-sight velocity
of component B has little gradient in the equatorial plane of the
region, but this might result in part from a combination of flow
divergence and the value of $B|{los}$ (which is measured close to the
centre of the region) being rather larger than that of $B_z$ in our
example.  While the qualitative properties of this assignment are
attractive, it remains to calculate a proper model tuned to the
properties of the region, in particular its geometry.  Nevertheless,
the present discussion at least illustrates how both fast and slow
shocks should be considered in the analysis of regions with
well-ordered magnetization.

In future work, we will model in detail the global structure of
magnetized \HII\ regions and the local structure of photoevaporated
magnetized clumps.

\subsection*{Acknowledgements.}

We thank Sam Falle and Serguei Komissarov for helpful discussions on
evolutionary conditions, and the referee for constructive comments
which brought several issues into sharper focus.  RJRW acknowledges
support from PPARC for this work.

\label{lastpage}
\end{document}